\documentclass[prd,nofootinbib,showpacs,preprint]{revtex4}

\newcommand{\be}{\begin{equation}}
\newcommand{\ee}{\end{equation}}
\newcommand{\bea}{\begin{eqnarray}}
\newcommand{\eea}{\end{eqnarray}}
\newcommand{\N}{\mathcal{N}}

\newcommand{\comment}[1]{}

\newcommand{\lb}{\label}

\newcommand{\alp}{\alpha'}

\begin{document}

\title{Graceful exit from a stringy landscape via MSSM inflation}

\author{Rouzbeh Allahverdi}
\email{rallahverdi@perimeterinstitute.ca}
\affiliation{Perimeter Institute for Theoretical Physics, Waterloo, ON, N2L
2Y5, Canada}

\author{Andrew R. Frey}
\email{frey@hep.physics.mcgill.ca}
\affiliation{Physics
Department, McGill University, Montr\'eal, QC, H3A 2T8, Canada}

\author{Anupam Mazumdar}
\email{anupamm@nordita.dk}
\affiliation{NORDITA, Blegdamsvej-17, Copenhagen-2100, Denmark}

\begin{abstract} 
The cosmological evolution of the string landscape is
expected to consist of multiple stages of old inflation with large
cosmological constant ending by tunnelling. Old inflation has a
well known graceful exit problem as the observable universe becomes
empty, devoid of any entropy. Simultaneously,
in the quest for reheating the right degrees
of freedom, it is important that the final stage of inflation reheat 
Standard Model sector.  It is known that inflation can occur naturally
along a flat direction of the Minimal Supersymmetric Standard Model (MSSM),
solving the reheating problem, but the initial conditions require
a large degree of fine-tuning.  In
this paper, we study how inflation of a MSSM flat direction can be
embedded into the string theory landscape to solve both the graceful
exit problem of old inflation and the fine-tuning problem of MSSM inflation, 
elaborating on ideas of Bousso and Polchinski.
The fluctuations of the MSSM flat direction during old inflation create regions
with initial conditions favorable for eternal inflation, which also 
allows the cosmological constant to continue to relax.  This final phase of
inflation also provides all the usual benefits of MSSM inflation, 
including straightforward reheating into Standard Model degrees of freedom.
\end{abstract}

\preprint{hep-th/yymmnnn}


\maketitle

\section{Introduction}

We expect that the early universe, prior to inflation, existed in a
state of high energy density, in particular, near the cut-off scale of
the 4D effective field theory we know today.  In string theory, this
cut-off is typically expected to be near the fundamental string scale.
Also, there is some evidence that string theory has a ``landscape'' of
metastable vacua with varying cosmological constant, moduli VEVs (and
therefore couplings), supersymmetry (SUSY) breaking scale, and so on,
which can be studied using statistical arguments (see
\cite{hep-th/0409207,hep-th/0601053,hep-th/0610102} for reviews).

Since many of these metastable vacua have large energy densities (as
we describe below), it seems likely that the early universe could have
existed as a de Sitter spacetime with a large cosmological constant,
which then decayed by tunnelling, as in old inflation
\cite{Guth:1980zm}.  In fact, such a cosmology with multiple stages
of inflation \cite{Burgess:2005sb} provides a mechanism by which the
full landscape of vacua is populated, as in 
\cite{Abbott:1984qf,hep-th/0004134,hep-th/0408133}.
With some caveats,
this mechanism can also relax the cosmological constant quickly
\cite{Abbott:1984qf,hep-th/0004134,hep-th/0408133,hep-th/0611148,
hep-th/0612056}.  In particular, the cosmological constant must be
able to decay even though other sectors will dominate once the energy
density reaches $\sim 1\, (\textnormal{TeV})^4$.

One obvious worry about this picture is that, if the universe were to
tunnel out of the false vacuum, then the universe would be devoid of
any entropy as the nucleated bubble would keep expanding forever with a
negative spatial curvature.  Such a universe would have no
place in a real world, so it is important that the last stage of
inflation be driven by a slow roll phase that must explain:
\begin{itemize}
\item{The observed temperature anisotropy in the cosmic microwave
background radiation and the subtle observational
details \cite{Spergel:2006hy}. This implies that the last tunnelling
event must occur at least 60 e-foldings before the end of inflation.}
\item{Reheating into Standard Model (SM) like degrees of freedom,
so that baryons and cold dark matter can be produced.}
\item To solve the cosmological constant problem, the slow-roll phase
must allow the population of vacua with smaller cosmological constants,
which is facilitated by eternal inflation.
\end{itemize}
The first criterion has been addressed by \cite{hep-th/0004134}, and we
will see that the Minimal Supersymmetric Standard Model (MSSM) can
facilitate all the criteria. 

Also, it was recently realized that MSSM has
all the necessary ingredients for successful inflation, whose
parameters can possibly be tested at the LHC
\cite{Allahverdi:2006iq,Allahverdi:2006cx,Allahverdi:2006we,Allahverdi:2006ng}.
Moreover, this inflation has a graceful exit which creates conditions
for producing baryons and cold dark matter.  The inflaton is a
\textit{gauge invariant} MSSM flat direction (for a review on MSSM
flat directions, see \cite{Enqvist:2003gh,Dine:2003ax}). There are
potentially two candidates, $LLe$ and $udd$, where the superfield $L$
stands for a $SU(2)_{L}$ lepton doublet, while $e,~u,~d$ are superfields
standing for the right-handed components of the electron and up and
down type quarks. Both the inflaton candidates carry SM lepton and
baryon numbers. After inflation, the inflaton decay directly creates baryons,
leptons, and cold dark matter (the lightest superpartner is a candidate
for cold dark matter \cite{Jungman:1995df}).\footnote{Often the
inflaton is treated as an absolute gauge singlet
\cite{hep-th/0503203}.  Even the string inspired models which often
use moduli, string axion and the dilaton to be the inflaton are all
absolute SM gauge singlets. In all these cases reheating into the SM
like degrees of freedom is not so obvious and plagued with the
uncertainty due to the inflaton coupling to the matter field
\cite{hep-th/0508139}. As an absolute gauge singlet the inflaton can have
any coupling strength to the SM degrees of freedom, there is no
symmetry argument which prohibits the coupling strength to be large or
small \cite{Allahverdi:2006wh}.}  This decay addresses the second point 
above.

Inflation within the MSSM occurs near a saddle point of the flat
direction potential, which is lifted by the soft SUSY breaking mass
term, the non-renormalizable superpotential, a potential supergravity
mass correction, and the $A$-term \cite{Dine:1995uk,Dine:1995kz}. All
these contributions play important role. Although none of the
individual terms would give rise to inflation \cite{Jokinen:2004bp},
all the terms combined provide a unique saddle point condition in
gravity mediated SUSY breaking case
\cite{Allahverdi:2006iq,Allahverdi:2006cx,Allahverdi:2006we} and in
gauge mediated SUSY breaking case \cite{Allahverdi:2006ng}. In these
cases, the first nontrivial derivative of the inflaton potential is
the third derivative, which alone determines the dynamics.  This MSSM
inflation matches the current WMAP observations with the right
amplitude and the spectral tilt
\cite{Allahverdi:2006we,Allahverdi:2006wt,BuenoSanchez:2006xk}.  One
important result is that the MSSM inflation produces many e-foldings
of slow roll inflation, $\mathcal{N}_{e}\sim 10^{3}$, with a preceding
self-reproduction regime.  Therefore, MSSM inflation can satisfy the
first and third criteria above.  One difficulty of MSSM inflation, though, 
is that it requires a $10^9$ fine-tuning of the initial conditions for 
slow-roll (and $10^{13}$ for self-reproduction) and
a $10^9$ finetuning of the $A$-term \cite{hep-ph/0610069,hep-ph/0610134}.

As we shall see, old inflation on the landscape and MSSM inflation 
complement each other nicely.  The landscape picture 
suggests that the $A$-term fine-tuning is no longer a problem.
Also, as in \cite{hep-th/0004134}, fluctuations
of the MSSM inflaton during false vacuum inflation naturally fine tunes
the initial conditions for MSSM inflation,  i.e. why the MSSM
inflation occurs near the saddle point of the potential
\cite{Allahverdi:2006iq,Allahverdi:2006we,Lyth:2006ec}.\footnote{This
would be difficult to explain otherwise. For example, one may not
invoke thermal effects to trap the flat direction near the saddle
point as it does not correspond to a false minimum, or a point of
enhanced symmetry.}  On the
flip side, MSSM inflation provides a period of eternal inflation followed
by many e-foldings of slow-roll inflation, which allows tunnelling to 
solve the cosmological constant problem and further creates the necessary
temperature anisotropies.  Finally, MSSM inflation, by its nature, allows
simple reheating into the SM sector.

We will begin our discussion with a broad brush-stroke review of the
string landscape and the SM-like sector of it, 
first showing why we expect the landscape can accomplish
fine-tuning of the $A$-term.  Then we discuss the time scales associated 
with tunnelling events on the landscape, showing why eternal inflation
seems necessary for the cosmological constant problem and why the
arguments of \cite{hep-th/0004134} require a slight extension.  We will
also show why we can assume the MSSM to be present throughout the chain
of tunnelling events.  Following our discussion of the MSSM in the 
landscape, we give a brief review of MSSM inflation.  Then we explain how
old inflation on the landscape can set the initial conditions for
eternal inflation in the MSSM sector, using an improved form of the
arguments of \cite{hep-th/0004134}.  In short, the MSSM flat
directions, being light, obtain random quantum jumps, which 
displace them to large expectation values \cite{hep-th/0503203}.  
Alternately, supergravity corrections to the MSSM
can trap the flat directions in a false minimum (as in
\cite{hep-ph/9503259,hep-ph/9503303,hep-ph/9507453}).

\section{Old Inflation on the Landscape}\label{s:landscape}

There are three main points we wish to make clear in our review of the
landscape picture of string theory vacua.  First, we wish to explain 
the current picture of landscape, including large (string scale)
cosmological constants, especially reviewing the distribution of SM-like
metastable vacua on the landscape.  We will also review the scale of 
SUSY breaking and 
corrections to the MSSM Lagrangian due to the landscape sector and argue
that the $A$-term can scan over the necessary range for MSSM inflation.
Second, we will discuss
tunnelling rates for the decay of each false vacuum, describing the
phase of old inflation prior to MSSM inflation.  With these rates in
hand, we will see why the original solution of \cite{hep-th/0004134} 
for the empty universe problem is not quite complete.  Finally, we will see
why eternal inflation is necessary for tunnelling on the landscape
to solve the cosmological constant problem in a realistic cosmology.

\subsection{Landscape with large $\Lambda$}\label{s:largecc}

The most basic fact about the landscape (which is usually not discussed,
being of little interest for present-day physics) is that the cosmological
constant is generically large.
A simple way to see that 
follows \cite{hep-th/0004134}, which describes the landscape
contribution to the cosmological constant as arising from string
theory flux.  In this picture, the vacuum energy
\be\lb{bp}
V=M_P^2\Lambda = M_P^2\Lambda_0 +\alp{}^{-2}\sum_i c_i n_i^2\ ,
\ee
where $c_i\lesssim 1$ are constants and $n_i$ are flux quantum
numbers (note that $\Lambda$ has dimension $\mathrm{mass}^2$
in our notation as it enters the Einstein equation as $\Lambda g_{\mu\nu}$).
It is clear that large $\Lambda$ corresponds to a large
radius shell in the space of flux quanta, so larger $\Lambda$ will
have more possible states.  All in all, string theory (from the
landscape point of view) could have from $10^{500}$ to even
$10^{1000}$ vacua
\cite{hep-th/0409207,hep-th/0601053,hep-th/0610102,hep-th/0004134},
with the vast majority of those having large cosmological constants.

In addition, our knowledge of the distribution of gauge groups over
the landscape suggests that one out of $10^{10}$ vacua have the SM
spectrum, at least in simple models
\cite{hep-th/0411173,hep-th/0510170,hep-th/0512190,hep-th/0606109}.
Even if this fraction is much smaller on the whole landscape, there
are so many vacua that it seems likely very many will have a SM-like
spectrum.  For simplicity's sake, we will consider only this SM-like
subset of the landscape in the rest of the paper, meaning we are
ignoring possible transitions among vacua with different gauge groups
in our cosmological model.  We will justify this assumption in section
\ref{s:allpaths}.

Given that the cosmological constant is large, at what scale is
supersymmetry in the SM sector?  We need to know this because we will want
to follow the behavior of MSSM degrees of freedom throughout our
cosmology, including time spent in vacua of large $\Lambda$, and we need
to estimate corrections to the MSSM action.  From
statistical arguments (again, see
\cite{hep-th/0409207,hep-th/0601053,hep-th/0610102} for more
references), most vacua should have badly broken supersymmetry, with
large F-terms in the supersymmetry breaking sector.\footnote{This holds
true even with small $\Lambda$.}  However, we assume that these
F-terms are not in the SM sector, which can therefore be described as
the MSSM (though perhaps with large soft breaking terms).  In
addition, there exist vacua with large cosmological constant that are
``almost supersymmetric'' in the sense of \cite{hep-th/0601053}.
These vacua have vanishing (or nearly vanishing) F-terms, and their
cosmological constant and supersymmetry breaking are provided by a
D-term, such as that created by an anti-brane.  Upon decay of the
D-term, the remaining vacuum has small $\Lambda$ and low energy
supersymmetry.  The original model of \cite{hep-th/0301240} is almost
supersymmetric in this sense because anti-D3-branes provide both the
cosmological constant and supersymmetry breaking.  We can also
consider the toy ``friendly landscape'' models of
\cite{hep-th/0501082}, in which the dynamics of $N$ scalars
create a landscape of vacua.  In the supersymmetric case,
\be\lb{friendly}
M_P^2\Lambda = M_{sb}^4-\frac{3|W|^2}{M_P^2}\ ,\ee
where $M_{sb}$ is the supersymmetry breaking scale of a hidden sector
(\textit{not} the $N$ scalars) and the superpotential falls in the
range
\be\lb{friendlyW}
0\leq |W|^2 \leq \sqrt{N} M_\star^6\ee
with $M_\star$ the cut-off.

In any of these cases, though, the scale of supersymmetry breaking in the
MSSM sector is controlled in the early universe by the form of
mediation.  If supersymmetry breaking is gravity mediated, then the
soft masses for the MSSM can be quite small $m^2\sim M_{sb}^4/M_P^2$,
while the vacuum energy ranges over the landscape from zero (or
potentially negative values) up to $M_{sb}^4$.  Even the very
non-supersymmetric vacua with large $\Lambda$ might be viable for our
purposes.  Suppose that the MSSM feels supersymmetry breaking from
both a landscape sector and a hidden sector (the hidden sector provides 
the present-day SUSY breaking), with gravity mediation
from the landscape sector.  At large $\Lambda$, we find that the
corrections to the MSSM soft masses from the landscape factor are
$m\sim \Lambda^{1/2}\sim H$, so these corrections are the same as the
Hubble corrections of
\cite{hep-ph/9503259,hep-ph/9503303,hep-ph/9507453}.  We will consider
these corrections explicitly below in section \ref{s:mssminflate}.

We also want to know whether we can expect the $A$ parameter of the MSSM
to vary over different values on the landscape in order to carry out the
fine-tuning of one part in $10^9$ necessary for MSSM inflation.
As it turns out, the $A$ term breaks a $U(1)$ symmetry of the flat direction
in the MSSM,\footnote{This can simply be an R symmetry that multiplies the
flat direction by a phase, which we can require the MSSM Lagrangian to 
respect before adding $A$ terms.} 
so the landscape should scan values of $A/M_P$ 
(the appropriate dimensionless combination) around $A=0$
in the sense of \cite{hep-th/0501082}.  In this case, we need a value of
$A\sim 10^{-16} M_P$ with a tuning of $\delta A/A\sim 10^{-9}$, so we need
at least $10^{25}$ MSSM vacua to achieve the right vacuum spacing.  Given
the large number of MSSM vacua that seem likely to exist, it seems a safe
assumption that the landscape can accomplish the necessary tuning of $A$.
Again, most of these SM-like or MSSM-like vacua will have
large $\Lambda$, so it seems reasonable to think that a generic
initial condition for cosmology is one of these large $\Lambda$ vacua.


\subsection{Decay time scales and inflation}\label{s:decaytime}

The decay rate per volume (by tunnelling) of a metastable vacuum to the
nearest neighboring vacuum\footnote{To avoid subtleties, we require that
the both states have nonnegative
energy density.  See, for example, \cite{hep-th/0211160}
for issues in the negative $\Lambda$ case.  Also, \cite{hep-th/0701083}
has discussed the importance of negative $\Lambda$ vacua in possibly
separating parts of the landscape from each other.  We adopt the view,
as discussed in that paper, that the landscape is sufficiently complicated
that there are no isolated regions.} takes the form
\be\lb{tunnel}
\Gamma/V = C \exp\left(-\Delta S_E\right)\ ,\ee
where $C$ is a one-loop determinant and $\Delta S_E$ is the difference
in Euclidean actions between the instanton and the background with
larger cosmological constant.  The determinant $C$ can at most be
$C\lesssim M_P^4$, simply because $M_P$ is the largest scale
available, and estimates (ignoring metric fluctuations) give a value
as small as $C\sim r^{-4}$, with $r$ the instanton bubble radius
\cite{hep-ph/9308280}.  If we therefore look at a decay rate in a
(comoving) Hubble volume, we find
\be\lb{tunnel2}
\Gamma\lesssim \frac{M_P^4}{H^3}\exp\left(-\Delta S_E\right)\,.
\ee
Especially with a large Hubble scale, the associated decay time is
much longer than $1/H$, given that typically $\Delta S_E\gg 1$.

In fact, as given in \cite{hep-th/0305018} following
\cite{Coleman:1980aw,Brown:1987dd,Brown:1988kg},
the Euclidean action takes the form
\be\lb{action}
\Delta S_E = 2\pi^2 r^3 \tau_e \,,
\ee
where the bubble radius and effective tension $\tau_e$ are given in
terms of dimensionless cosmological constants
$\lambda=M_P^4\Lambda/\tau^2$, and the actual tension of the bubble
wall $\tau$.  The full formulae are listed in the appendix.

Unfortunately, a given tunnelling process in the landscape probably has
$\lambda_+\sim 1$ and $\lambda_-\lesssim 1$.  However, it may be
instructive to look at three limits, $\lambda_+\to 0,\infty$.  In the
latter case, we simultaneously take $\lambda_-\to 0$ or
$\lambda_-\to\lambda_+$ (these are the same for
$\lambda_+\to 0$).  As $\lambda_+$ vanishes, we find
\be\lb{mink}
\Delta S_E\to 24\pi^2 \frac{M_P^6}{\tau^2\lambda_+}=24\pi^2
\frac{M_P^2}{\Lambda_+}\, .
\ee
(The tension surprisingly drops out!)
For $\lambda_+\to\infty$, we find the limits
\bea
\Delta S_E&\to& 12\pi^2\frac{M_P^6}{\tau^2\lambda_+}=
12\pi^2\frac{M_P^2}{\Lambda_+}\ \ (\lambda_-\to 0)\ ,\lb{drop}\\
\Delta S_E&\to& 6\sqrt{3}\pi^2\frac{M_P^6}{\tau^2\lambda_+^{3/2}}
=6\sqrt{3}\pi^2\frac{\tau}{\Lambda_+^{3/2}}\ \ (\lambda_-\to\lambda_+)\, .
\lb{step}
\eea
The only parametrically different behavior is the last limit as
$\lambda_{\pm}\to\infty$, which means either that the tension is very
small or that the cosmological constants are both very large (or that
$M_P\to\infty$, corresponding to field theory without gravity).
Technically, the instanton approximation can break down in the limit
(\ref{step}), if the Euclidean action becomes much smaller than one
(as a full quantum mechanical treatment would be necessary), but the
important point is that $\Gamma/H\gtrsim 1$ if that is the case.  From
(\ref{tunnel2}), we might expect that $\Gamma/H\gg 1$, but we should
remember that the coefficient of the exponential in (\ref{tunnel2}) is
an upper limit and that quantum mechanical effects could limit the
decay rate to $\Gamma \sim H$.

Therefore, we expect small jumps in $\Lambda$ with small tension
$\tau$ to be the most common decays, which can in fact be quite rapid.
Recently,
\cite{hep-ph/0412145,hep-ph/0502177,hep-th/0612056} have argued that
a series of decays in this last limit can give a novel form of
inflation (dubbed ``chain inflation'') in which each metastable vacuum
gives less than an e-folding, but which can overall have many
e-foldings.\footnote{Recently, \cite{hep-th/0612222} has discussed the
set of vacua in a particular region of the landscape, and this
``topography'' may be relevant for discussions of chain inflation.}
In particular, in a large volume flux compactification
model (ignoring warping) like those of \cite{hep-th/0004134} there is
a range of flux quanta such that $\Delta S_E\ll 1$ (and $\Lambda
<M_P^2$), so $\Gamma/H\sim 1$.  In a warped flux compactification, the
limit (\ref{step}) can still apply, but, using the logic of
\cite{hep-th/0112197,hep-th/0305018,hep-th/0508139},
the tunnelling rate is somewhat slower, but similar (see the appendix
for details of these calculations).  However, in the landscape, chain
inflation does not have an \textit{a priori} graceful exit
\cite{hep-th/0612056}; the argument of \cite{hep-th/0004134} does not quite
apply because it requires a large change in cosmological constant in the
final decay before low-scale inflation (without inflaton 
trapping).  In section \ref{s:mssminflate}, we give an improved form of the
argument of \cite{hep-th/0004134} that shows how old inflation on the
landscape can source MSSM inflation without inflaton trapping and with only 
small jumps in the cosmological constant.

Note that, when MSSM inflation starts, the ``bare'' cosmological
constant (that not associated with the MSSM inflaton) might still be
considerably larger than the present value.  This means that further
instanton decays should take place to reduce the bare cosmological
constant, and these decays should occur during MSSM inflation in order
for the bubble regions to grow long enough.  
Since it seems likely that the instanton
bubble tension will be large compared to the scale of MSSM inflation,
the decay rate will be given by (\ref{mink}), which is highly
suppressed.  This would then require MSSM inflation to last for
extremely many e-foldings.  Fortunately, MSSM inflation naturally
includes a self-reproduction (eternal inflation) regime prior to
slow-roll \cite{Allahverdi:2006iq,Allahverdi:2006cx,
Allahverdi:2006we}.  In other words, it seems that a low-scale eternal 
inflation is a necessary part of using decays to solve the cosmological
constant problem, and MSSM inflation includes it naturally.
Eternal inflation solves several of problems relating
false vacuum inflation to observations, and it is particularly viable
at the TeV scale of MSSM inflation \cite{hep-th/9805167,hep-th/0011262}.

We have two more comments regarding additional types of transitions on
the landscape. First, \cite{hep-th/0611148} argues that resonance can play
an important role in tunnelling across a landscape of many metastable
vacua; we would not find qualitative differences in that case, but
decays could occur more quickly.  Second, we have
discussed thin-wall instantons, as were first studied in
\cite{Coleman:1980aw,Brown:1987dd,Brown:1988kg}.  However, in some
cases of very large tension, so-called Hawking-Moss instantons
\cite{Hawking:1981fz} can dominate the decay rate.  This typically
occurs when the potential barrier separating two vacua is very wide,
but, unless the barrier is also very shallow, the decay rate is
approximately given by (\ref{mink}).  Because the potential barriers
are typically expected to be quite tall on the landscape (up to
string-scale energy densities), we will not treat Hawking-Moss
instantons separately. Note also that there has been some recent
discussion in the literature on the physical meaning of the
Hawking-Moss instanton (e.g. \cite{hep-th/0612083,hep-th/0612146}),
but we will not delve into those subtleties here.

\subsection{Following all paths}\label{s:allpaths}

As has been emphasized by \cite{hep-th/0004134,hep-th/0408133}, the whole
landscape, including the MSSM-like vacua, will be populated eventually
almost independently of initial conditions due to repeated transitions
among the different states on the landscape.  In fact, if we ignore sinks
in the landscape, eventually all the states will be populated with a fraction
of the comoving volume given by a thermal distribution (see 
\cite{hep-th/0701083,hep-th/0605266,hep-th/0611043} for a recent discussion
including sinks).  

What is important for us is that, because transitions are possible to states
with higher cosmological constant, each bubble of a given state will contain
bubbles of every other state.  This fact means that any given path
through the vacua of the landscape will be followed by some region of
comoving volume.  In particular, we can consider a history in which some
patch of comoving coordinates enters a state with an MSSM sector with
the appropriate tuning of the $A$-term at high cosmological constant and
then proceeds to decay by reducing the cosmological constant without affecting
the MSSM sector in any significant way.  Therefore, we are justified in 
assuming that we can follow the behavior of MSSM fields and their actions
throughout the process of old inflation on the landscape.  Incidentally,
these considerations are necessary for any application of the arguments of
\cite{hep-th/0004134} for the exit from old inflation; the low energy
inflaton must exist in the states with larger cosmological constant.


\section{MSSM inflation}\label{s:mssminflate}

Let us now discuss the what happens to the observable sector in the
background of inflation spacetime. As we have noticed, the
universe cascades from large cosmological constant to
another somewhat smaller as time progresses.
Given the decay time, it is
fairly evident that huge number of e-foldings can be generated.

However, in order to have a sensible phenomenology, it is important
that the last phase of inflation is primarily driven by MSSM-like
vacua. Furthermore, it is equally pertinent that there exists a
graceful exit of inflation within MSSM. This is possible provided
there is no potential barrier in the MSSM like sector. Otherwise, the
universe would nucleate in a cold state with a negative spatial
curvature, because the bubble collision rate would not overcome the
expansion rate \cite{Guth:1980zm}.  In this regard, slow roll
inflation in the MSSM sector as discussed in
\cite{Allahverdi:2006iq,Allahverdi:2006cx,Allahverdi:2006we,Allahverdi:2006ng}
is ideal.  We now ask how the dynamics of the MSSM flat
direction will unfold during the era of false vacuum inflation.

During the false vacuum inflation the energy density of the universe,
and hence the expansion rate $H_{\mathrm{false}}$, remains constant
for a given $\Lambda$. The flat direction potential receives
corrections from soft SUSY breaking mass term, the non-renormalizable
superpotential correction, $W\sim \lambda\Phi^{n}/M_{P}^{n-3}$, where
$W$ is the superpotential, $\Phi$ denotes the flat direction and
$\lambda_{n}$ an $\mathcal{O}(1)$ coefficient,\footnote{$\lambda_n$
might actually be larger or smaller; for example, the cut-off might be
given at a smaller scale (such as a Kaluza-Klein scale) or $\lambda_n$
might be suppressed as in \cite{hep-th/9502069}.  However, $\lambda_n$
changes only the VEV of $\Phi$ and not the physics of inflation, so we
will ignore this point.}  we also assume that the cut-off is given by
the $4$-dimensional Planck mass,\footnote{The non-renormalizable term
can be barred if the flat direction under consideration preserves
R-symmetry; otherwise, they are lifted by higher order superpotential
operators, see
\cite{Gherghetta:1995dv,Dine:1995uk,Dine:1995kz}.} and corrections
due to the large Hubble expansion for a minimal choice of K\"ahler
potential. These corrections, parameterized by two constants $c\sim
\mathcal{O}(1)$ and $a\sim \mathcal{O}(1)$, result in
\cite{Dine:1995uk,Dine:1995kz,Enqvist:2003gh,Dine:2003ax}\footnote{The
origin of the Hubble induced terms is due to couplings between the modulus
which drives the false vacuum inflation and the MSSM sector. Apriori,
even in string theory, the K\"ahler potential for the modulus is not
well known at all points of the parameter space. Similarly, within the MSSM,
the K\"ahler potential is unknown. Therefore, the coefficients $c,a$
are not fixed. For a no-scale type model, the Hubble induced
corrections are vanishing at tree level; however, they do appear at
one loop $| c | \sim 10^{-2}$
\cite{Gaillard:1995az,Allahverdi:2001is}.}
\be \label{hubblepot}
V = \frac{1}{2} (m^2_{\phi} + c H^2_{\mathrm{false}}) | \phi |^2 +
\left[(A + a
H_{\mathrm{false}}) \lambda_n \frac{\phi^n}{n M^{n-3}_{P}} +
{\mathrm{h.c.}})\right] + \lambda^2_n \frac{|\phi |^{2(n-1)}}
{M^{2(n-3)}_{P}}\,,
\ee
where the soft SUSY breaking mass term is generically small compared
to the Hubble expansion rate of the false vacuum, $m_{\phi}\sim 1\
\mathrm{ TeV}\ll H_{\mathrm{false}}$. We define $\Phi\equiv \phi
e^{in\theta}$, and the dynamics of $\phi$ in an inflationary
background depends on $c$ and $a$.  Therefore we consider different
cases separately.

It turns out that the dynamics of the MSSM flat direction largely follow the
physics discussed in \cite{hep-th/0004134}.  We review and
elaborate on their discussion in the context of the MSSM and give an 
important improvement on their result in the context of negligible Hubble
corrections.


\subsection{Positive Hubble induced corrections}

A positive Hubble induced correction provides $c\sim +\mathcal{
O}(1),a\sim \mathcal{O}(1)$. This is a typical scenario when the
K\"ahler potential for the string modulus comes with a canonical
kinetic term. Although this could be treated as a special point on the
K\"ahler manifold, it is nevertheless important to discuss this
situation. This is also the simplest scenario out of all
possibilities. A generic flat direction gets a large mass of the order
of Hubble expansion rate; therefore, its fluctuations are unable to
displace the flat direction from its global minimum.

On phenomenological grounds, this is an uninteresting and undesirable
case, since the bubble does not excite any of the MSSM
fields. Therefore, the bubble remains empty and devoid of energy with
no graceful exit of inflation from the false vacuum. The universe
continues cascading to smaller $\Lambda$ with smaller
$H_{\mathrm{false}}$.  Eventually, when $H_{\mathrm{false}}\leq
m_{\phi}$, the MSSM flat directions would be free to move. However,
through this time the fields were never displaced from their minimum
and therefore the dynamics of the flat directions would remain
frozen. The universe would be cold, as it was before, and the spatial
curvature would remain negative. It is fair to say that, on
phenomenological grounds, such a universe is already ruled out.
This paves the way for more interesting scenarios, which we discuss next.


\subsection{Negligible Hubble induced corrections}

In this case, the potential is not affected by the false vacuum
inflation at all, namely $|c|,|a|\ll 1$.  So long as $V^{\prime
\prime}(\phi) \ll H^2_{\mathrm{false}}$, the flat direction field
$\phi$ makes a quantum jump of length $H_{\mathrm{false}}/2\pi$ within
each Hubble time.\footnote{To be more precise, the quantum
fluctuations of $\phi$ have a Gaussian distribution, and the r.m.s.~of
modes which exit the horizon within one Hubble time is
$H_{\mathrm{false}}/2 \pi$.} These jumps superpose in random walk
fashion resulting in \cite{hep-th/0503203}
\be \label{diff}
\Big(\frac{d \langle \phi^2 \rangle}{dt}\Big)_{fluctuations} =
\frac{H^3_{\mathrm{false}}}{4 \pi^2}.
\ee
On the other hand, the classical slow roll due to the potential leads to
\be \label{slow}
\Big(\frac{d \langle \phi^2 \rangle}{dt}\Big)_{slow~roll} = - \frac
{2 \langle V^{\prime}(\phi) \phi \rangle}{3 H_{\mathrm{false}}}.
\ee
For a massive scalar field $V(\phi)\sim m^2_{\phi} \phi^2/2$, the combined
effects yield \cite{hep-th/0503203}
\be \label{fluct}
\langle \phi^2 \rangle = \frac{3 H^4_{\mathrm{false}}}{8 \pi^2 m^2_{\phi}}
\left[1 - \exp\left(-\frac{2 m^2_{\phi}}{3 H_{\mathrm{false}}}
t\right)\right]\,.
\ee
The maximum field value
\be \label{rms}
\phi_{r.m.s.} = \sqrt{\frac{3}{8 \pi^2}} \frac{H^2_{\mathrm{false}}}{m_{\phi}},
\ee
at which the slow roll motion (\ref{slow}) counterbalances the random
walk motion (\ref{diff}), is reached for $\Delta t
\gg 3H_{\mathrm{false}}/2 m^2_{\phi}$. This amounts to a number
\be
\label{falsefold}
\mathcal{N}_{\mathrm{false}} \gg \frac{3}{2}
\left(\frac{H_{\mathrm{false}}}{m_{\phi}}\right)^2.
\ee
of e-foldings of inflation in the false vacuum.

In the absence of Hubble induced corrections, the potential
in~(\ref{hubblepot}) has a saddle point at
\be \label{saddle}
\phi_{0}=\left(\frac{m_{\phi}M_{P}^{n-3}}
{\lambda_{n}\sqrt{2n-2}}\right)^{1/(n-2)}\,,
\ee
where $V^{\prime}(\phi_0)=V^{\prime\prime}(\phi_0)=0$ while
$V^{\prime\prime\prime}(\phi_0)\neq 0$, provided that
$A^2=8(n-1)m_{\phi}^2$.\footnote{This can happen in the gravity
mediated case, where $A \sim m_{\phi} \sim m_{3/2}$, where
$m_{3/2}\sim \mathcal{O}(1~\mathrm{TeV})$ is the gravitino mass. The
situation is quite different in the case of a gauge mediated SUSY
breaking scenario \cite{Allahverdi:2006ng}.}  For $m_{\phi}\sim
100~\mathrm{GeV}-10$ TeV and $\lambda_n \sim {\cal O}(1)$, and for $n=6$,
the VEV is $\phi_0\sim 10^{14}-10^{15}$ GeV. The suitable flat
directions are $LLe$ and $udd$, which are lifted by $n=6$
superpotential terms and also have a non-zero $A$-term as required by
the condition for a saddle
point \cite{Allahverdi:2006iq,Allahverdi:2006we}.\footnote{In order to
have successful inflation, we require the condition
$A^2=8(n-1)m_{\phi}^2$ to be satisfied to one part in
$10^{9}$. Although this requires a fine tuning, SUSY can allow it to
be maintained order by order if $A/m_{\phi}$ acts as an infrared fixed
point of the renormalization group
flow \cite{Allahverdi:2006we}. Also, as noted in section
\ref{s:largecc} above, this tuning can be explained naturally by the
landscape picture.  For larger deviations there is a point of
inflection with large $V^{\prime}(\phi_0)$ (or a negligible $A$-term
discussed in \cite{Jokinen:2004bp}), or a pocket of false minimum
\cite{Allahverdi:2006dr}. Neither case leads to a slow roll inflation
within the MSSM.}

For $\phi < \phi_0$ the mass term dominates the flat direction potential.
Then quantum fluctuations can push $\phi$ to the vicinity of
$\phi_0$ if $\phi_{r.m.s} \geq \phi_0$.\footnote{Due to the Gaussian
distribution of fluctuations, the probability of having $\phi \gg
\phi_{r.m.s}$ is exponentially suppressed.} This, according to (\ref{rms}),
requires that
\be \label{cond1}
H_{\mathrm{false}} \geq \left(\frac{8 \pi^2}{3}\right)^{1/4} (m_{\phi}
\phi_0)^{1/2} \simeq 10^{9}~\mathrm{GeV}.
\ee
The number of e-foldings needed for this to happen is
\be \label{efold}
\N_{\mathrm{false}} \leq \left(\frac{H_{\mathrm{false}}}{10^9~\mathrm{GeV}}
\right)^2 10^{12}.
\ee
Indeed for $H_{\mathrm{false}} \geq \phi_0$ the inherent uncertainty due to
quantum fluctuations implies that $\phi > \phi_0$ within one Hubble time.

A last stage of MSSM inflation with an expansion rate
\be \label{hinf}
H_{\mathrm{MSSM}} =\frac{n-2}{\sqrt{6n(n-1)}}\frac{m_{\phi}\phi_{0}}
{M_{P}} \sim {\cal O}(1~{\mathrm{GeV}})
\ee
starts if $V(\phi)$ dominates the energy density of the universe,
i.e. $V(\phi) > 3 H^2_{\mathrm{false}} M^2_{P}$. An
observationally consistent inflation in the slow roll regime requires that the
displacement from the saddle point satisfy
$| \phi - \phi_0 | < \Delta \phi$,
where~\cite{Allahverdi:2006iq,Allahverdi:2006we}
\be \label{domain}
\Delta \phi = \frac{\phi^3_0}{4n(n-1) M^2_{P}} \simeq
10^6~\mathrm{GeV}.
\ee
In the landscape, the universe can begin in a false vacuum with
arbitrarily large $H_{\mathrm{false}}$ (as long as
$H_{\mathrm{false}} \ll M_{P}$). Therefore, we generically expect
that $\phi$ is quickly pushed to field values $\phi \gg \phi_0$.
However, $H_{\mathrm{false}}$ slowly decreases as a result of
tunnelling to vacua with smaller cosmological constant~\footnote{Note
that we need to stay in an MSSM-like vacuum all the way until MSSM
inflation begins. Given the scarcity of MSSM-like vacua in the
landscape, the probability of tunnelling from one such vacuum to
another is $\sim 10^{-9}$.
However, due to eternal inflation in the false vacuum, the physical
volume of the universe increases by a factor of ${\rm exp}(3 H_{\rm
false}/\Gamma)$ within a typical time scale for bubble nucleation
($\Gamma$ is the false vacuum decay rate, see subsection 2.2). This
easily wins over the suppression factor $10^{-9}$ for $\Gamma < 9
H_{\rm false}$.}. For a massive scalar field with the potential
$V(\phi)\sim m^2_{\phi} \phi^2/2$, this implies a gradual decrease
of $\phi_{r.m.s}$, see (\ref{rms}). Since quantum fluctuations can
at most push $\phi$ to $\phi_{r.m.s}$, this also implies that $\phi$
is slowly decreasing in time.  Indeed for $H_{\mathrm{false}} <
10^9$ GeV, we find that $\phi < \phi_0$, irrespective of how large
$\phi$ initially was.  This is the case in the discussion of 
\cite{hep-th/0004134}, which is why \cite{hep-th/0004134} requires a jump
from large cosmological constant directly to the slow-roll inflationary
stage.

Note, however, that the potential becomes very flat around $\phi_0$ as a
result of the interplay among different terms in
(\ref{hubblepot}). In fact, for $| \phi - \phi_0 | \ll
\phi_0$ we have \cite{Allahverdi:2006iq,Allahverdi:2006we}
\be \label{3rd}
V(\phi) \approx V(\phi_0) + \frac{1}{3(n-2)^2}
\frac{m^2_{\phi}}{\phi_0} (\phi - \phi_0)^3\,.
\ee
Once $H_{\mathrm{false}} \sim 10^9$ GeV, $\phi$ reaches this plateau (from
above).  Within the plateau, $V^{\prime}(\phi)$ becomes increasingly
negligible, and so does the classical slow roll; they exactly vanish
at $\phi = \phi_0$. In consequence, quantum jumps dominate the dynamics
and freely move $\phi$ throughout the plateau. Hence the flatness of
potential, which is required for a successful MSSM inflation,
guarantees that $\phi$ will remain within the plateau during the
landscape evolutionary phase.  It is possible to generalize this argument
to other models of low-scale inflation; we see that a sufficiently flat
inflaton potential can trap inflaton fluctuations in the slow-roll region.

The flat direction eventually dominates the energy density of the
universe when $H_{\mathrm{false}} < 1$ GeV, see (\ref{hinf}). MSSM
inflation then starts in those parts of the universe which obey
(\ref{domain}). Having $\phi$ so close to $\phi_0$ is possible
since the uncertainty due to quantum fluctuations is $\sim 1$ GeV
at this time.

The false vacuum inflation paves the way for an MSSM inflation inside
the nucleated bubble.  The modulus
which was responsible for the
false vacuum inflation continues tunnelling to minima
with smaller (eventually the currently observed)
cosmological constant inside the MSSM inflating bubble. The
modulus
will oscillate around the minimum of its potential as the
curvature of its potential dominates over the Hubble expansion rate. The
fate of oscillations would depend on the coupling of the modulus to
the MSSM fields. For a coupling of gravitational strength,
oscillations are long-lived. Moreover, the decay will be kinematically
forbidden if the decay products are coupled to the flat direction
$\phi$ (which has obtained a large VEV $\simeq \phi_0$). However, such
details are largely irrelevant as the flat direction dominates the energy
density of the universe and drives inflation, diluting the energy density in
oscillations.

In particular, we note that MSSM inflation has a
self-reproducing regime \cite{Allahverdi:2006iq,Allahverdi:2006we}
because $V^{\prime}(\phi)$ is extremely small and the potential becomes
very flat close to $\phi_0$.  The observable part of our universe
can spend an arbitrarily long time in the self-reproducting regime
before moving into the standard slow-roll inflation.  During this
period, the cosmological constant can continue to decay and eventually
settle at an observationally acceptable value.

It is already known from \cite{Allahverdi:2006iq,Allahverdi:2006we}
that the flat direction generates a large number of e-foldings during
slow roll inflation, i.e. $\mathcal{N}_{e}\sim 10^{3}$, and during
this phase the amplitude of the scalar perturbations and the spectral index are
given by \cite{Allahverdi:2006iq,Allahverdi:2006we}
\bea\label{saddle0}
\delta_{H} &\equiv &\frac{1}{5\pi}\frac{H^2_{\mathrm{MSSM}}}{\dot\phi}\simeq
\frac{1}{5\pi}\sqrt{\frac{1}{6}n(n-1)(n-2)}\left(
\frac{m_{\phi}M_{P}}{\phi_0^2}\right)\mathcal{N}_{\mathrm{COBE}}^2\,,\\
\label{saddle1}
n_s &=&1+2\eta -6\epsilon \simeq 1-\frac{4}{\mathcal{N}_{\mathrm{COBE}}}\,.
\eea
Here $\mathcal{N}_{\mathrm{COBE}}\sim 50$ is the relevant number of
e-foldings of inflation required to explain the current
observations. For $m_{\phi}\sim 100~{\mathrm{GeV}}-10$ TeV, $n=6$, $\phi_0\sim
10^{14}-10^{15}$ GeV, we obtain the correct amplitude of density
perturbations and the scalar spectral index to be $n_s=0.92$
within $2\sigma$ of the current WMAP results.\footnote{The above
results, (\ref{saddle0},\ref{saddle1}) are strictly true in the saddle
point limit. A small deviation from the saddle point condition
predicts a spectral tilt $0.92\leq n_s\leq 1.0$ with an observable
range and negligible running and tensor perturbations
\cite{BuenoSanchez:2006xk,Allahverdi:2006wt}.}  Actually, the value of
$\mathcal{N}_{\mathrm{COBE}}$ depends on the thermal history of the
Universe. In our case, the precise thermal history is known because
the MSSM flat direction has SM couplings, and we know the maximum (see
below) and the reheat temperatures
\cite{Allahverdi:2006wh,Allahverdi:2006xh,Allahverdi:2005fq,Allahverdi:2005mz}.

Slow roll inflation ends when the slow roll conditions are
violated. Eventually, the flat direction starts oscillating around the origin
with a frequency $m_{\phi}\sim 1$ TeV. The flat
direction condensate dcays via instant preheating
and creates a relativistic bath of MSSM
quanta with a maximum temperature
\cite{Allahverdi:2006iq,Allahverdi:2006we}
\be
T_{max}\sim [V(\phi_0)]^{1/4}\sim 10^{9}\ \mathrm{GeV}\,.
\ee
The temperature is sufficiently high to trigger electroweak
baryogenesis within the MSSM and to provide the sufficient thermal
condition for production of cold dark matter, i.e. the
Lightest Supersymmetric Particle (LSP).

Before we conclude this discussion, let us remind the readers that
obtaining the flat direction VEV, $\phi_0$, depends on the false
vacuum inflation, which requires $H_{\mathrm{false}}\geq 10^{9}$ GeV
at some time
(see (\ref{cond1})).  This condition, although very probable in the landscape
picture, need not be satisfied
always. Then the quantum fluctuations would not be large enough to push the
flat direction to the vicinity of $\phi_0$ as required for a final stage of
MSSM inflation. This would therefore ruin the inflationary and phenomenological
predictions. This can be avoided if the flat
direction is trapped in a false minimum which evolves with time, as discussed
below.


\subsection{Negative Hubble induced corrections}

The case with $c \sim -\mathcal{O}(1)$ may arise naturally for
non-minimal K\"ahler potential~\cite{Dine:1995uk,Dine:1995kz}. For
$H_{\mathrm{false}} \gg m_{\phi}$ the potential in (\ref{hubblepot})
becomes tachyonic, and its true minimum is located at
\be \label{true}
\phi_{\mathrm{min}} \simeq \left(\frac{\sqrt{| c |}}{\lambda_n \sqrt{2n-2}}
H_{\mathrm{false}} M_{P}^{n-3}\right)^{1/(n-2)}\,.
\ee
Note that $\phi_{\mathrm{min}}$ is initially larger than $\phi_0$, see
(\ref{saddle}).

The curvature of the potential at the minimum is $V^{\prime \prime}
(\phi_{\mathrm{min}}) = (4n-7) | c | H^2_{\mathrm{false}}
\gg H^2_{\mathrm{false}}$.
This implies that $\phi$ is driven away from the origin, due
to quantum fluctuations, and quickly settles down at $\phi_{\mathrm{min}}$.
The MSSM flat direction is \textit{trapped} inside the minimum, held
due to false vacuum inflation, and gradually tracks the instantaneous
value of $\phi_{\mathrm{min}}$
as $H_{\mathrm{false}}$ decreases.\footnote{Here we
are assuming that the difference between the energy densities of the
two false vacua is small compared to their average energy density.}

The minima at $\phi = 0$ and $\phi = \phi_{\mathrm{min}}$ become degenerate
when $H_{\mathrm{false}} \sim m_{\phi}$. For $H_{\mathrm{false}} \ll m_{\phi}$
the true minimum is at $\phi = 0$ and the one at $\phi_{\mathrm{min}}$ will be
\textit{false}. The Hubble induced corrections are subdominant in this
case. We then find \cite{Allahverdi:2006we}
\begin{eqnarray}
& & \phi_{\mathrm{min}} \simeq \phi_0 \left(1 + \frac{1}{n-2}
\frac{\sqrt{| c |} H_{\mathrm{false}}}{m_{\phi}}\right)
\, \\ \label{falmin}
& & \, \nonumber \\
& & V^{\prime \prime}(\phi_{\mathrm{min}}) \simeq 2 (n-2)
\sqrt{| c |} H_{\mathrm{false}} m_\phi \, . \label{falcurv}
\end{eqnarray}
The $\phi$ field can track down the instantaneous
value of $\phi_{\mathrm{min}}$ provided that $\sqrt{V^{\prime
\prime}(\phi_{\mathrm{min}})}$ is greater than the Hubble expansion rate. In
fact this is the case so long as
$H_{\mathrm{false}} > H_{\mathrm{MSSM}} \sim 1$ GeV,
see (\ref{falcurv}). Once $H_{\mathrm{false}} \simeq H_{\mathrm{MSSM}}$,
the flat
direction potential dominates the energy density of the universe,
and MSSM inflation
begins at a Hubble expansion rate $H_{\mathrm{MSSM}}$.
In the meantime, landscape
tunnelling to vacua with smaller
cosmological constant continues, and the location of the false minimum
$\phi_{\mathrm{min}}$
continuously changes.\footnote{Note that tunnellings do not
affect the Hubble expansion rate anymore since the flat direction dominates
the energy density of the universe now.} Eventually $V^{\prime
\prime}(\phi_{\mathrm{min}}) < H^2_{\mathrm{MSSM}}$ when
\be \label{end}
H_{\mathrm{false}} \simeq \frac{1}{2(n-2)}
\frac{H^2_{\mathrm{MSSM}}}{m_{\phi}},
\ee
at which time
\be \label{falminend}
\phi_{\mathrm{min}} - \phi_0 \simeq \frac{\phi_0}{2(n-2)^2} \left(
\frac{H_{\mathrm{MSSM}}}{m_{\phi}}\right)^2.
\ee
It turns out from (\ref{hinf},\ref{domain}) that $\phi_{\mathrm{min}} -
\phi_0 \ll \Delta \phi$.  Therefore $\phi$ is already inside the
interval required for a successful MSSM inflation. At this point the
Hubble induced corrections become largely unimportant, and all the
successes of MSSM inflation are retained, as discussed in the previous
subsection. The fate of the string moduli inside the MSSM bubble
remains the same as in the previous subsection.  In particular, it does not
matter whether the
universe tunnels right away to the currently observed value of
$\Lambda$ or not. Inflation dilutes any
excitations of the modulus oscillations during inflation.  Our Hubble
patch reheats when the MSSM flat direction rolls down to its minimum
and starts creating MSSM quanta as discussed in the previous
subsection.

Note that all needed for the success of this scenario is to start in a false
vacuum in the landscape
where $H_{\mathrm{false}} > m_{\phi} \sim 1$ TeV. This ensures that the flat
direction will roll way from the origin and settle at
$\phi_{\mathrm{min}}$, which
is the true minimum of $V(\phi)$ at that
time. It will then track $\phi_{\mathrm{min}}$
as $H_{\mathrm{false}}$ slowly decreases,
and will eventually land inside the appropriate interval around $\phi_0$.
This is a much milder condition than that in the case of negligible Hubble
induced corrections $H_{\mathrm{false}} \geq 10^9$ GeV (see the previous
subsection).

One comment is in order. This scenario has some similarities to the
Affleck-Dine baryogenesis with negative Hubble induced
corrections \cite{Dine:1995uk,Dine:1995kz}. It can be seen from
(\ref{hubblepot}) that the equation of motion of the flat
direction has a fixed point so long as Hubble induced corrections are
dominant. The flat direction tracks the fixed point in a radiation or
matter dominated universe.  However, a non-adiabatic change in the
potential occurs when $H \sim m_{\phi}$, and the flat direction starts
oscillating around the origin.  However, in our case the universe is
in a de Sitter phase with a slowly varying $H_{\mathrm{false}}$, and the
flat direction tracks the false minimum of its potential until it
dominates the universe. As explained, this is necessary for having a
successful MSSM inflation.


\section{Conclusion and Discussion}

In this paper, we have developed a modified version of ``chain
inflation'' which has a graceful exit due to MSSM inflation.  Multiple
stages of false vacuum inflation relax the cosmological constant and
provide the appropriate initial conditions for inflation driven by the
MSSM flat direction \cite{Allahverdi:2006iq,Allahverdi:2006we} (as follows
from an improved version of arguments from \cite{hep-th/0004134}).  This
last stage inflates the universe to its current size and generates
acceptable density perturbations which match the current
observations.  Furthermore, MSSM inflation will dilute any relics of string
theory and has a neglible gravity wave background, so it has a clear
prediction of standard inflationary CMB perturbations with no gravity
waves, which is relevant for coming experiments.
Without the initial stages of high energy false vacuum
inflation, it would be very difficult to explain the initial conditions
for MSSM inflation, and without the latter there would be no graceful
exit from the false vacuum inflation.  We find the complementarity of
these two types of inflation to be intriguing.

Moreover, MSSM inflation produces a self-reproduction regime and many
e-foldings of slow-roll inflation, which is important in solving the
cosmological constant problem via tunnelling.  As discussed in
\cite{hep-th/9805167,hep-th/0011262}, self-reproduction allows the
cosmological constant to settle down to its present value long before the
final 60 e-foldings of inflation, which allows for a standard CMB
perturbation spectrum.  Also, \cite{hep-th/0011262} points out that TeV
scale eternal inflation can fit observational constraints, even after
tunnelling events in the context of string and M theory, which is another
point of interest for MSSM inflation.  In addition, self-reproduction is
a virtue which is difficult to find
among inflationary mechanism at scales below the string scale. For
instance, brane---anti-brane inflation (for a recent review see
\cite{Cline:2006hu}) and multiple-axion-driven
\cite{Dimopoulos:2005ac} assisted inflation
\cite{Liddle:1998jc,Copeland:1999cs} fail to provide large e-foldings
of inflation without fine-tuning of the Lagrangian.  On the other
hand, ``roulette inflation'' \cite{hep-th/0612197} and ``racetrack
inflation'' \cite{hep-th/0406230,hep-th/0603129} can give rise to
self-reproduction, and it would be interesting to study either of
these models as a graceful exit from false vacuum inflation on the
landscape and how false vacuum inflation can provide the proper
initial conditions for those types of inflation.


\acknowledgments

We are thankful to Raphael Bousso, Robert Brandenberger, Kari Enqvist, and
Juan Garcia-Bellido
for discussion at various stages of this work. The work of RA
is supported by Perimeter Institute for Theoretical Physics. Research
at Perimeter Institute is supported in part by the Government of
Canada through NSERC and by the province of Ontario through MRI.  ARF
is partly supported by an IPP/PI postdoctoral fellowship and partly by
le Fonds Nature et Technologies du Qu\'ebec. AM is partly supported by
the European Union through Marie Curie Research and Training Network
``UNIVERSENET'' (MRTN-CT-035863).

\appendix
\section{Tunnelling formulae}

For completeness, we list here the instanton bubble radius and effective
tension:
\be\lb{radius}
\frac{1}{r^2} = \frac{\tau^2}{M_P^4} \left[ \frac{1}{16}+
\frac{1}{6}(\lambda_+ +\lambda_-) +\frac{1}{9}
(\lambda_+-\lambda_-)^2\right]\,,
\ee
and
\bea \lb{tension}
\tau_e &=& \tau\left\{1+\frac{6}{\lambda_+\lambda_-}\left[
\lambda_+\left|\frac{1}{4}+\frac{1}{3}(\lambda_+-\lambda_-)\right|^3
+\lambda_-\left|\frac{1}{4}+\frac{1}{3}(\lambda_--\lambda_+)\right|^3
\right.\right.\nonumber\\
&&\left.\left.
+(\lambda_--\lambda_+)\left(\frac{1}{16}+\frac{1}{6}(\lambda_++\lambda_-)
+\frac{1}{9}(\lambda_+-\lambda_-)^2\right)^{3/2}\right]\right\}\ .
\eea

We will also give the details of the decay rate calculation for the
two examples of chain inflation listed in section \ref{s:decaytime}.
The first case, ignoring warping, was given in \cite{hep-th/0612056},
and we will recap here.  For a perturbative large volume
compactification with many fluxes, the cosmological constant is given
by (\ref{bp}) with the constants $c_i\ll 1$.  We also take an
isotropic compactification, so all the $c_i$ are approximately equal.
Further, the brane charge $q$ and Planck mass (we follow here the
conventions of \cite{hep-th/0612056} except that we take $q\to M_P^2
q$) become
\be\lb{mpq} q\approx c\ , \ \ M_P^2\approx 1/c\alp\ .  \ee
The instanton bubble tension is then $\tau\sim M_P/\alp$, so we find
$\lambda_+\sim M_P^2\alp c n^2\sim n^2\gg 1$.  Finally, the Euclidean
action of the instanton is
\be\lb{dsechain} \Delta S_E \approx
\frac{\tau}{\Lambda_+^{3/2}}\approx \frac{1}{cn^3}\ .\ee
As long as $c^{-1/3}\ll n\ll c^{-1}$, $\Delta S_E\ll 1$ and
$\Lambda_+\ll M_P^2$.

In the second case, we look at a compactification with a similar
5-brane instanton decay localized in a warped throat.  This decay has
been studied in \cite{hep-th/0112197,hep-th/0305018}.  A perturbative
$g_s\sim 1/10\lesssim 1$ large volume compactification $e^u\sim 10
\gtrsim 1$ (with $e^u$ the linear scale of the (unwarped)
compactification manifold) has a Newton constant
\be\lb{newton}
\frac{1}{2\kappa_4^2}\approx \frac{1}{2\pi\alp}\ .\ee
(Here, we follow the conventions of \cite{hep-th/0305018}, in which
the VEVs of $u$ and $g_s$ are scaled out of the Einstein frame
metric.)  It is possible to see that the instanton bubble tension is
dominated by a term
\be\lb{kpvtau} \tau\approx
\frac{1}{g_s^{1/2}e^{6u}} \frac{|z|}{\alp{}^{3/2}}\ ,\ee
where $z$ is the complex structure modulus of the warped throat.  This
modulus is related to the minimum warp factor of the throat in such a
way that
\be\lb{kpvtau2} \tau\approx \ell_{sw}^{-2} (g_sM)^{3/2}
g_s^{-5/4} e^{-9u} \approx 10^{-2} \ell_{sw}^{-2}\ ,\ee
where $M$ is a flux quantum number and $g_s M\sim 10\gtrsim 1$ and
$\ell_{sw}$ is the warped string length at the bottom of the throat.

Now let us imagine that this decay does not annihilate most of the
cosmological constant.  Then \cite{hep-th/0508139} argues that
$\ell_{sw}^{-2}\sim\Lambda$.  This immediately implies that
$\lambda_+\gtrsim M_P^4/\Lambda^2\gg 1$ and $\Delta S_E\sim 1$.  (We
know that $\lambda_-\sim \lambda_+$ because the energy density
localized in the warped region is $1/\ell_{sw}^4\sim \Lambda^2\ll
M_P^2\Lambda$.)

\bibliography{oldinf}

\end{document}